\documentclass[preprint]{aastex}

\usepackage{color}
\usepackage{ulem}

\usepackage{xspace}

\newcommand{\xmm}{\emph{XMM-Newton}\xspace}

\newcommand{\atca}{\emph{ATCA}\xspace}
\newcommand{\CTIO}{\emph{CTIO}\xspace}
\newcommand{\Nanten}{\emph{Nanten}\xspace}

\newcommand{\SII}{[S\,{\sc ii}]}

\newcommand{\OIII}{[O\,{\sc iii}]}
\newcommand{\Halpha}{H${\alpha}$}
\newcommand{\D}{$^\circ$}
\newcommand{\h}{\small$^\textrm{h}$}
\newcommand{\m}{\small$^\textrm{m}$}

\def\xmm{XMM-{\it Newton}}

\def\HI{\hbox{H\,{\sc i}}}
\def\HII{\hbox{H\,{\sc ii}}}
\def\p0{\phantom{0}}
\def\it{\sl}
\def\degr{\hbox{$^\circ$}}
\def\arcmin{\hbox{$^\prime$}}
\def\arcsec{\hbox{$^{\prime\prime}$}}
\def\SBN70{\mbox{{N\,70}}}

\shorttitle{Multi-frequency observations of LHA\,120--N\,70}
\shortauthors{A. Y. De Horta et al.}

\begin{document}


\title{Multi-frequency observations of  a superbubble in the LMC: The case of LHA\,120--N\,70}


\author{A.~Y.~De Horta, E.~R.~Sommer, M.~D.~Filipovi\'c, A.~O'Brien, L.~M.~Bozzetto, J.~D.~Collier, G.~F.~Wong, E.~J.~Crawford and N.~F.~H.~Tothill}
\affil{University of Western Sydney, Locked Bag 1797, Penrith South DC, NSW 1797, Australia}
\email{a.dehorta@uws.edu.au}

\and
\author{P.~Maggi and F.~Haberl}
\affil{Max-Planck-Institut f{\"u}r extraterrestrische Physik, Giessenbachstrasse, 85741 Garching, Germany}




\begin{abstract}
We present a detailed study of new Australia Telescope Compact Array (\atca) and \xmm\ observations of LHA\,120--N\,70 (hereafter \SBN70), a spherically shaped object in the Large Magellanic Cloud (LMC) classified as a superbubble (SB). Both archival and new observations were used to produce high quality radio-continuum, X-ray and optical images. The radio spectral index of \SBN70\ is estimated to be $\alpha$=--0.12$\pm$0.06 indicating that while a supernova or supernovae have occurred in the region at some time in the distant past, \SBN70\ is not the remnant of a single specific supernova. \SBN70\ exhibits limited polarisation with a maximum fractional polarisation of 9\% in a small area of the north west limb. We estimate the size of \SBN70\ to have a diameter of 104~pc ($\pm$1~pc). The morphology of \SBN70\ in X-rays closely follows that in radio and optical, with most X-ray emission confined within the bright shell seen at longer wavelengths. Purely thermal models adequately fit the soft X-ray spectrum which lacks harder emission (above 1~keV). We also examine the pressure output  of \SBN70\ where the values for the hot ($P_{\rm X}$) and warm ($P_{\rm \HII}$) phase are consistent with other studied \HII\ regions. However, the dust-processed radiation pressure ($P_{\rm IR}$) is significantly smaller than in any other  object studied in \citet{2013arXiv1309.5421L}. \SBN70\ is a very complex region that is likely to have had multiple factors contributing to both the origin and evolution of the entire region. 
\end{abstract}



\section{Introduction}
\label{s:intro}

The Large Magellanic Cloud (LMC), located at a distance of 50~kpc \citep{2013Natur.495...76P}, is one of the best galaxies in which to study objects such as (super)bubbles, \HII\ regions and supernova remnants (SNRs), due to its favourable position in the direction toward the South Ecliptic Pole. As well as its viewing position, the LMC is also located in one of the coldest areas of the radio sky, which allows us to observe radio emission without the interruption from Galactic foreground radiation. In addition to this, the LMC resides outside of the Galactic Plane and therefore the influence of dust, gas and stars is negligible. 

A well-known characteristic of SNRs in the radio-continuum is the predominance of non-thermal emission. Although SNRs have a typical radio spectral index of $\alpha\sim-0.5$ (defined by $S\propto\nu^\alpha$) indicating a non-thermal emission, this can significantly change due to environment the SNR evolves in \citep{1998AAS..130..421F}. 

A bubble structure can be created by the stellar wind emanating from a massive O or B type star. This stellar wind collides with the surrounding interstellar medium (ISM) producing a hollowed out structure with a shock front, leading to a shell like nebula \citep{1977ApJ...218..377W}. This bubble may become bigger and more complex with the combined stellar wind from multiple stars in a cluster. As the most massive of these stars begin to explode as supernovae (SNe) the combined effect produces a (super)bubble. The effects from stellar winds and SNe become even more complex when taking place within \HI\ or \HII\ regions and where the ISM is heterogeneous \citep{1996ApJ...467..666O, 2004ApJ...613..302O}. 

These luminous superbubbles (SB), measuring up to 100~parsecs, blown by winds from hot massive stars and supernova explosions, with their interior filled with hot expanding gas offer us a chance to explore the crucial connection between the life-cycles of stars and the evolution of galaxies.

Here, we report on new radio-continuum, X-ray and optical observations of the LMC superbubble (SB) called LHA\,120--N\,70 (hereafter \SBN70). The observations, data reduction and imaging techniques are described in Section~2. The astrophysical interpretation of newly obtained moderate-resolution total intensity and polarimetric images in combination with the existing Magellanic Cloud Emission Line Survey (MCELS) and X-ray \xmm\ images are discussed in Section~3.

\subsection{Previous studies of \SBN70}
\label{ss:prev obs}

 \subsubsection{Optical}

\SBN70\ was first catalogued as an emission nebula in the Magellanic Clouds by \cite{1956ApJS....2..315H} while \cite{1976MmRAS..81...89D} listed this complex region DEM\,L301 which also encompasses the OB association LH\,114 \citet{1970AJ.....75..171L}. This stellar  association (LH\,114) is positioned about 1\arcmin\ west of the geometric centre of \SBN70 and in the line of sight of a known molecular cloud \citep{2009ApJS..184....1K}. It contains six O~stars with an O3If the most massive of these (Zhang et al. 2014, submitted). The first optical spectroscopy study by \cite{1976PASP...88...44D} found strong \SII\ lines relative to \Halpha\ indicative of an SNR although the electron density appeared low. \cite{1981A&A....97..342R} and \cite{1983A&AS...54..459G} measured the expansion velocity of \SBN70\ to 60--70~km~s$^{-1}$ and concluded that this would fit a shock model. They also suggested that \SBN70\ is most likely an ancient SNR approximately 2.4~$\times$10$^5$~yr old, in the adiabatic or radiative phase of expansion and although the central stars are contributing to the ionisation, they are not the origin of the shock. \cite{1981ApJ...250..103D} found that the spectrophotometry of \SBN70\ is inconsistent with the emission from an ionising shock, yet the \SII\ lines were moderately strong compared to ordinary \HII\ regions suggesting enhanced ionisation due to compression from stellar wind. \cite{1990MNRAS.246..358I} used a number of criteria such as a central star, nature of nebulosity and filament structure to define \SBN70\ as a possible Wolf--Rayet stellar wind bubble. \cite{1996ApJS..104...71O,1996ApJ...467..666O} describe \SBN70\ as a high-velocity SB -- an \HII\ region with OB association stars found within, creating a bubble from a combination of stellar wind and supernovae. \cite{1997MNRAS.291..827O} conclude \SBN70\ is an \HII\ region with OB association stars and that the \HII\ region is density--bounded rather than ionisation bounded. \cite{2006ApJ...646..205D} detected [O\textsc{vi}] in Far Ultraviolet Spectroscopic Explorer (FUSE) observations of four large stars in \SBN70, possibly from thermal conduction at the interface between the hot, X-ray emitting gas from inside the bubble and the photoionised material of the outer shell seen in \Halpha. 

 \subsubsection{Radio}

\cite{1980MNRAS.191..469M} performed initial radio observations of \SBN70\ at 5~GHz and 14.7~GHz and estimated a spectral index of $\alpha=-0.3$. \cite{1981ApJ...250..103D} found that unlike previous studies declaring \SBN70\ a non-thermal source, the radio emissions from \SBN70\ are predominantly  thermal in origin. They also found that although \SBN70\ exhibits several features that could be explained in  terms of an SNR or a mass-loss bubble, that the only model that is consistent with all observations is that of  a mass-loss bubble confined by the ram pressure of a massive, collapsing, cloud of neutral-hydrogen. However, \cite{1984AuJPh..37..321M} declared \SBN70\ was likely a fossil SNR, whilst  \cite{1998PASA...15..128F}\ list \SBN70\ as an \HII\ region Based on  Parkes radio observations.

 \subsubsection{X-ray}

\cite{1990ApJ...365..510C} (also see \cite{1991ApJ...373..497W} and \cite{1997AJ....113.1815C}) measured diffuse X-ray emission of bubbles and SBs and concluded \SBN70\ was an SNR within a wind-blown bubble. \cite{2011ApJ...732...98S} describe \SBN70\ as a SB but found no difference between \HII\ regions and SBs in IR emission from dust. \cite{2011ApJ...733...34R} carried out a simulation exploring the morphology, dynamics and thermal X-ray emission of SBs and concluded the structure of \SBN70\ can be explained with both stellar driven wind and a SN. \cite{2011RMxAC..40..199R} discuss the \xmm\ observations and the soft X-ray emission detected within \SBN70\ but also present an image showing some peripheral hard X-ray emission. Most recently Zhang et al. (2014, submitted), on top of their \xmm\ analysis, obtained high-dispersion long-slit echelle spectroscopic observations. They found that although \SBN70\ has a modest expansion velocity, its diffuse X-ray emission of \mbox{$\sim6.1\times10^{35}$~erg~s$^{-1}$} is higher than the luminosity from a quiescent superbubble with \SBN70Õs density, size and expansion velocity.

\section{Observational Data}
\label{s:obs data}

\subsection{Radio-continuum observations}
\label{ss:radio}

The data used in this study to produce the radio images of \SBN70\ were gathered from the radio interferometer Australia Telescope Compact Array (\atca). We observed \SBN70\ on the 15$^\mathrm{th}$ and 16$^\mathrm{th}$ of November~2011 with the \atca, using the new Compact Array Broadband Backend (CABB) receiver with the EW367 array configuration at wavelengths of 3 and 6~cm ($\nu$=9000 and 5500~MHz). Baselines formed with the $6^\mathrm{th}$ \atca\ antenna were omitted, as the other five antennas were arranged in a compact configuration. The observations were carried out in the so called ``snap-shot'' mode, totaling $\sim$70 minutes of integration over a 14~hour period. PKS~B1934-638 was used for primary calibration and  PKS~B0530-727 was used for secondary (phase) calibration. The \textsc{miriad} \cite{2006Miriad} and \textsc{karma} \cite{1996ASPC..101...80G} software packages were used for reduction and analysis. More information on the observing procedure and other sources observed in this session/project can be found in \cite{2007MNRAS.378.1237B}, \cite{2008SerAJ.176...59C,2008SerAJ.177...61C,2010A&A...518A..35C}, \cite{2009SerAJ.179...55C}, \cite{2012A&A...540A..25D}, \cite{2012A&A...539A..15G}, \cite{2012A&A...546A.109M} and \cite{2010SerAJ.181...43B,2012SerAJ.184...69B,2012MNRAS.420.2588B,2012RMxAA..48...41B,2013MNRAS.432.2177B}.

In addition to our own observations, we made use of two other \atca\ projects (C354 \& C149) at wavelengths of 13 \& 20~cm. Observations from project C354 were taken on the 18$^\mathrm{th}$ (array 1.5B), 22$^\mathrm{nd}$ and 23$^\mathrm{rd}$ (array 1.5D) of September 1994. Observations from project C149 were taken on the 22$^\mathrm{nd}$ of March 1992 (array 6A) and the 2$^\mathrm{nd}$ of April 1992 (array 6C).

Tables \ref{tbl:1} and \ref{tbl:2} summarise the data used for imaging in this paper. Table~\ref{tbl:1} includes archival observations from 1997 as well as our new observations made in 2011. In Table~\ref{tbl:2} we list archival data that were retrieved in the raw, uncalibrated form and were calibrated and used to produce the 20~cm images.

\begin{figure*}
 \includegraphics[angle=-90,width=0.5\columnwidth]{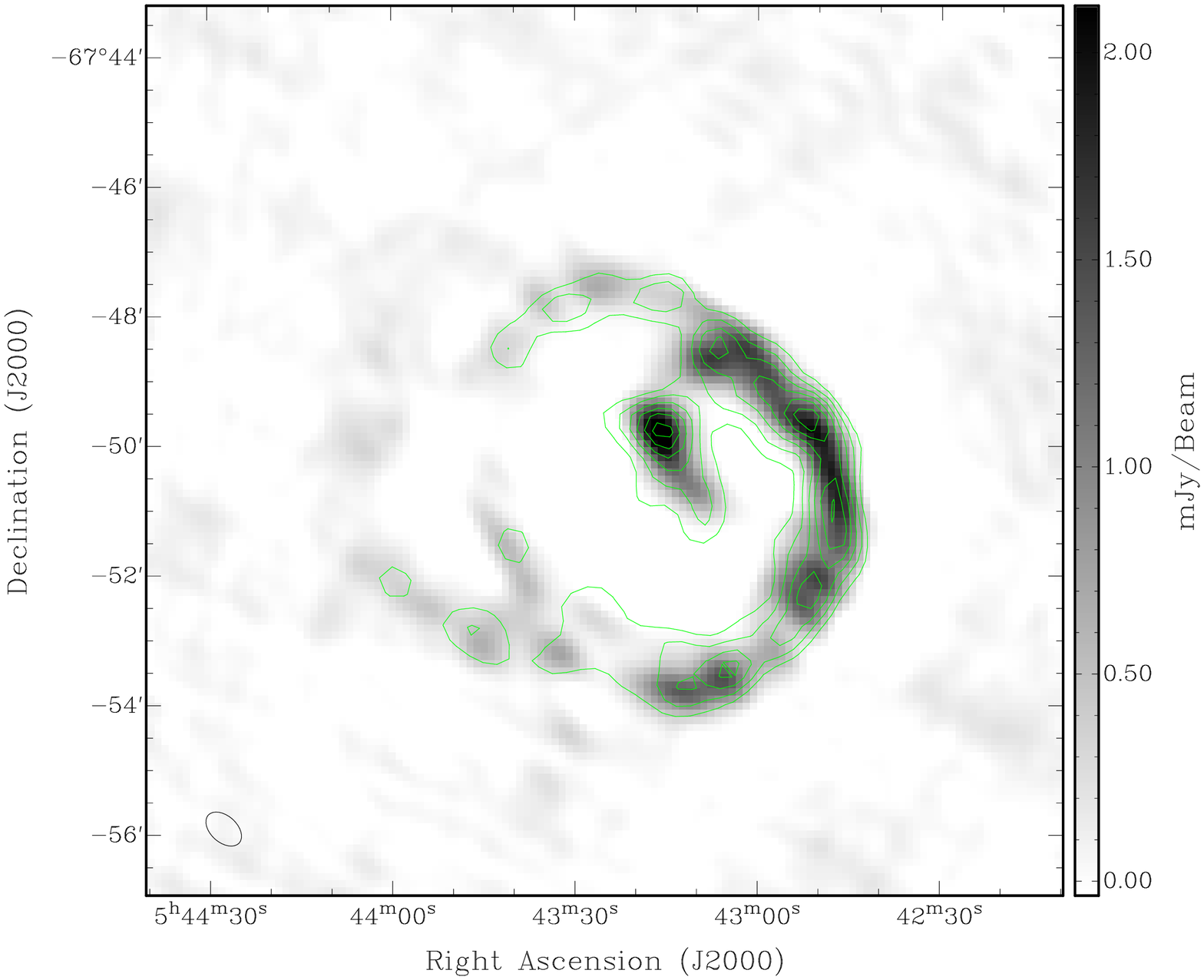}~\includegraphics[angle=-90,width=0.5\columnwidth]{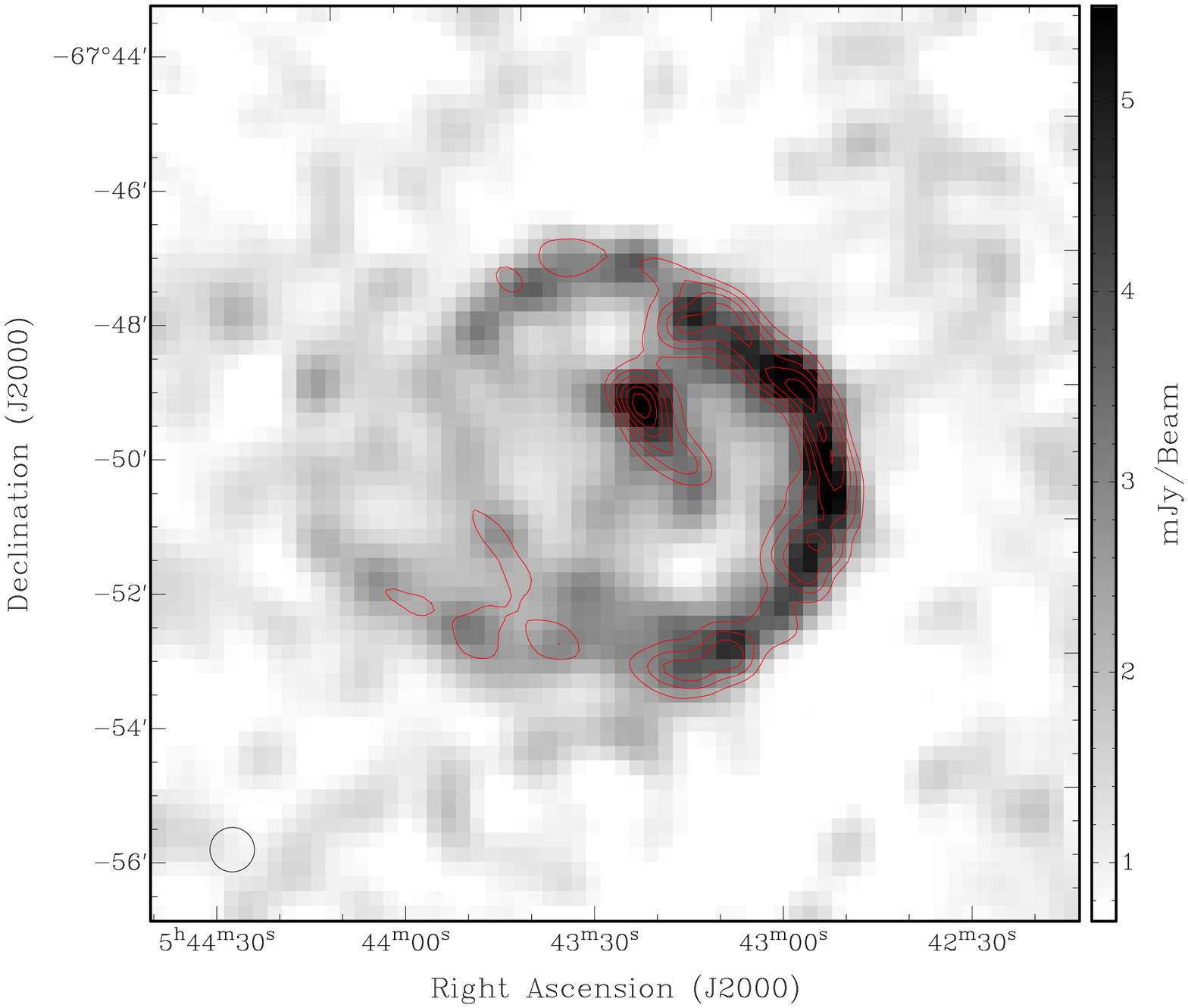}
 \caption{Left: ATCA observations of \SBN70\ at 6~cm (5.5~GHz) overlaid with 20~cm (1.4~GHz) contours at 3, 4, 5, 6 and 7$\sigma$ ($\sigma$=0.85~mJy~beam$^{-1}$). The black circle in the lower left corner represents the synthesised beamwidth (at 6~cm) of 39\,\arcsec$\times$25\arcsec at PA=48$^{\circ}$. 
Right: ATCA observations of \SBN70\ at 20~cm (1.4~GHz) overlaid with 6~cm (5.5~GHz) contours. The contours are 3, 6, 9, 12, 15 and 18$\sigma$ ($\sigma$~=~0.12~mJy~beam$^{-1}$). The black circle in the lower left corner represents the synthesised beamwidth (at 20~cm) of 40\,\arcsec$\times$40\arcsec. The sidebars quantifies the pixel map and its units are mJy~beam$^{-1}$.} 
 \label{fig:1}
\end{figure*}

Images were formed using \textsc{miriad}s multi-frequency synthesis algorithm \citep{1994A&AS..108..585S}  and natural weighting. They were deconvolved using the {\sc mfclean} and {\sc restor} algorithms with primary beam correction applied using the {\sc linmos} task. A similar procedure was used for both \textit{U} and \textit{Q} Stokes parameter maps. Because of the low dynamic range self-calibration was not attempted. The 6~cm image (Figure~\ref{fig:1}; left) has a resolution of 39\arcsec$\times$25\arcsec\ at PA=48\D\ with an estimated r.m.s. noise of 0.12~mJy~beam$^{-1}$. Similarly, we made an image of \SBN70\ at 20~cm (Figure~\ref{fig:1} right) with resolution of 40\arcsec$\times$40\arcsec\ at PA=0\D\ with an estimated r.m.s. noise of 0.85~mJy~beam$^{-1}$. 

 \subsection{\xmm\ observations}
  \label{ss:xmm}

We make use of the only existing \xmm\ observation of \SBN70\ (ObsId: 0503680201, PI: Rosa Williams). The 39 ks-long observation was performed on 2008-Jan-26, with \SBN70\ at the focus point of the European Photon Imaging Camera (EPIC). The EPIC instrument comprises a pn CCD imaging camera \citep{2001A&A...365L..18S} and two MOS CCD imaging cameras \citep{2001A&A...365L..27T}. The ``thin'' optical filter was used for all three cameras, which were operated in full-frame mode. We used the \xmm\ SAS\,\footnote{Science Analysis Software, \url{http://xmm.esac.esa.int/sas/}} version 12.0.1 to process the data. Periods of high background activity were screened out from the imaging and spectral analysis. We did so by applying a threshold of 8 and 2.5~cts\,ks$^{-1}$\,arcmin$^{-2}$ on pn and MOS background light curves in the 7--15~keV energy band. We were left with a useful exposure time of $\sim 22$~ks.

\begin{figure*}
 \centerline{\includegraphics[angle=270,width=1\textwidth, clip=true, trim=0cm 0cm 0cm -1.5cm]{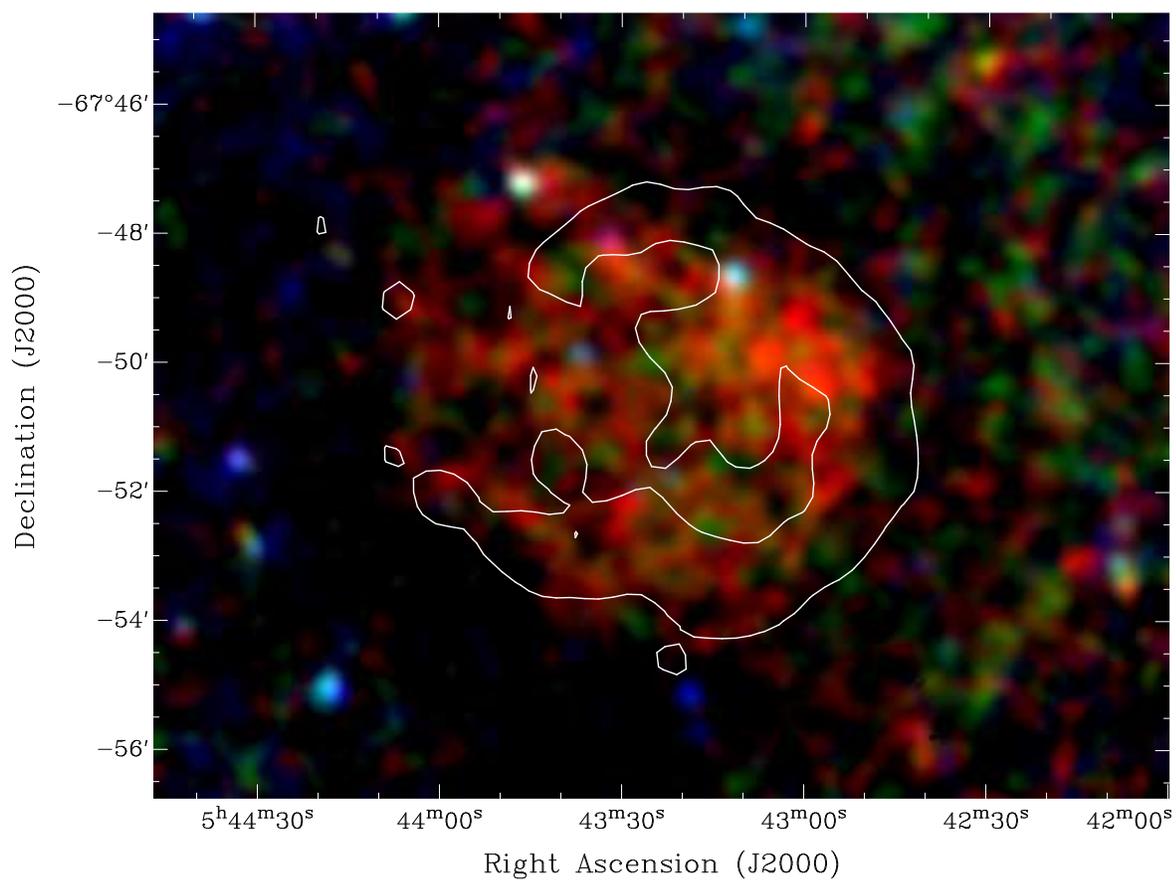}}
 \caption{\xmm\ X-ray (R = soft (0.3--0.7~keV), G = medium (0.7--1.1~keV), B = hard 1.1--4.2~keV)) image with 20~cm contour at 2.1~mJy~beam$^{-1}$.}
 \label{fig:2}
\end{figure*}

We created adaptively-smoothed, vignetting-corrected, and detector-background-subtracted images using the same method as described in \citet{2013MNRAS.432.2177B}. In Figure~\ref{fig:2} we combine \xmm\ images in three energy bands (a soft band (0.3--0.7~keV), a medium band (0.7--1.1~keV) and a hard band (1.1--4.2~keV)) into a composite and overlay radio (20~cm) contours.

For spectral analysis we utilised pn data only, as this camera has a higher throughput than the MOS cameras. We extracted energy spectra from a vignetting-weighted event list. The spectrum of \SBN70\ (``source spectrum'') was taken from a circle, whose centre and radius were as measured in radio (at 6~cm, Sect.\,\ref{s:resdis}). A background spectrum was extracted from a region of similar area located at the south-east of \SBN70\, excluding point sources detected in this region. Spectral fitting was performed with XSPEC \citep{1996ASPC..101...17A} version 12.8.0. Instead of subtracting the background spectrum from the source spectrum, we modelled the background and simultaneously fit both spectra. For a description of this method, as well as a detailed presentation of the instrumental and astrophysical components of the background, we refer the reader to \citet{2013MNRAS.432.2177B}.

 \subsection{MCELS}
  \label{ss:optical}
\begin{figure*}
 \centerline{\includegraphics[width=\textwidth]{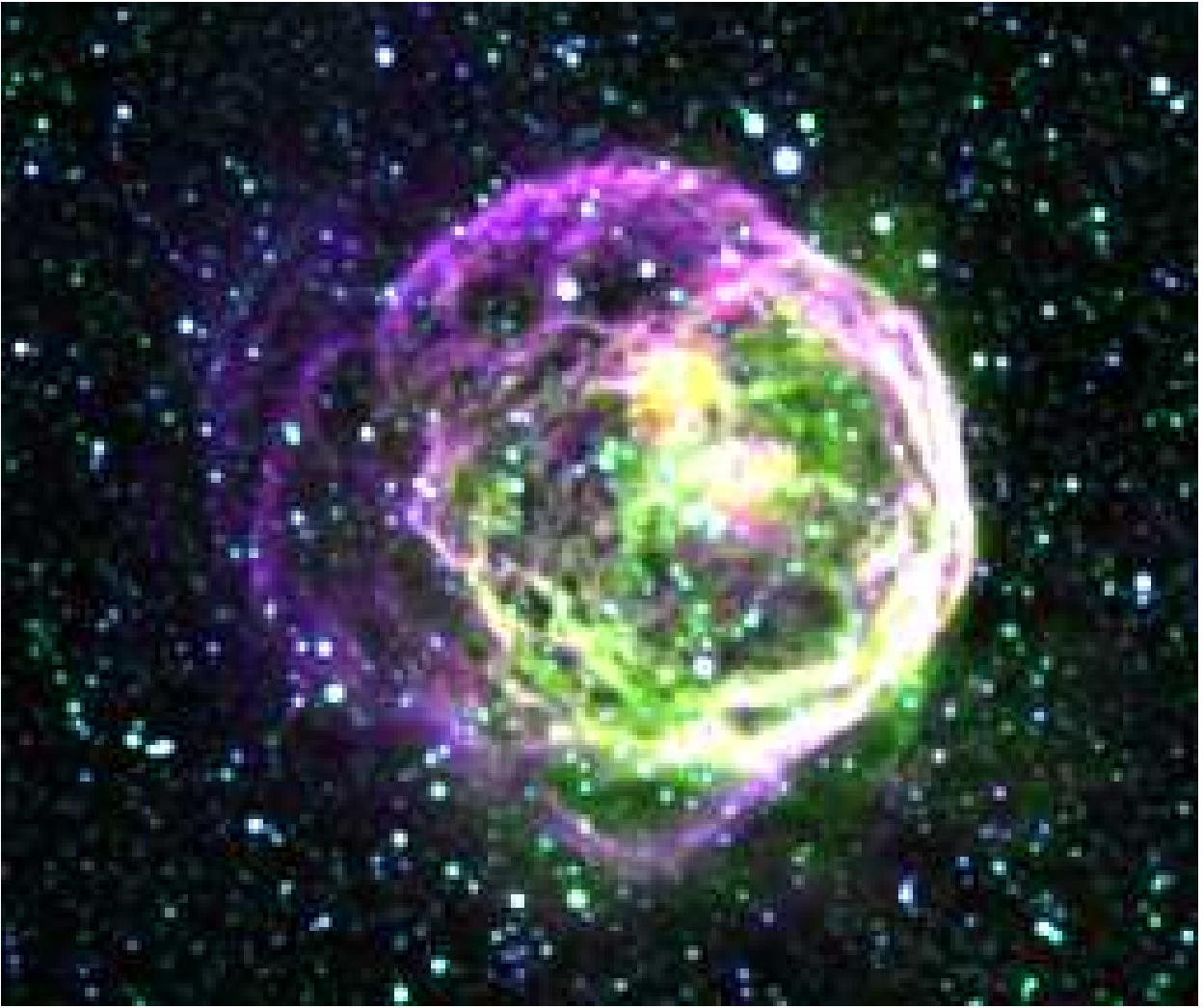}\vspace{-0.75 cm}\vspace{-0.65 cm}}
 \caption{The giant SB complex \SBN70\ in the light of \Halpha\ (red), \SII\ (green) and \OIII\ (blue); all data from MCELS (see Sect. 2.4).}
 \label{fig:3}
\end{figure*}

\begin{figure*}
 \centerline{\includegraphics[angle=-90,width=.33\textwidth]{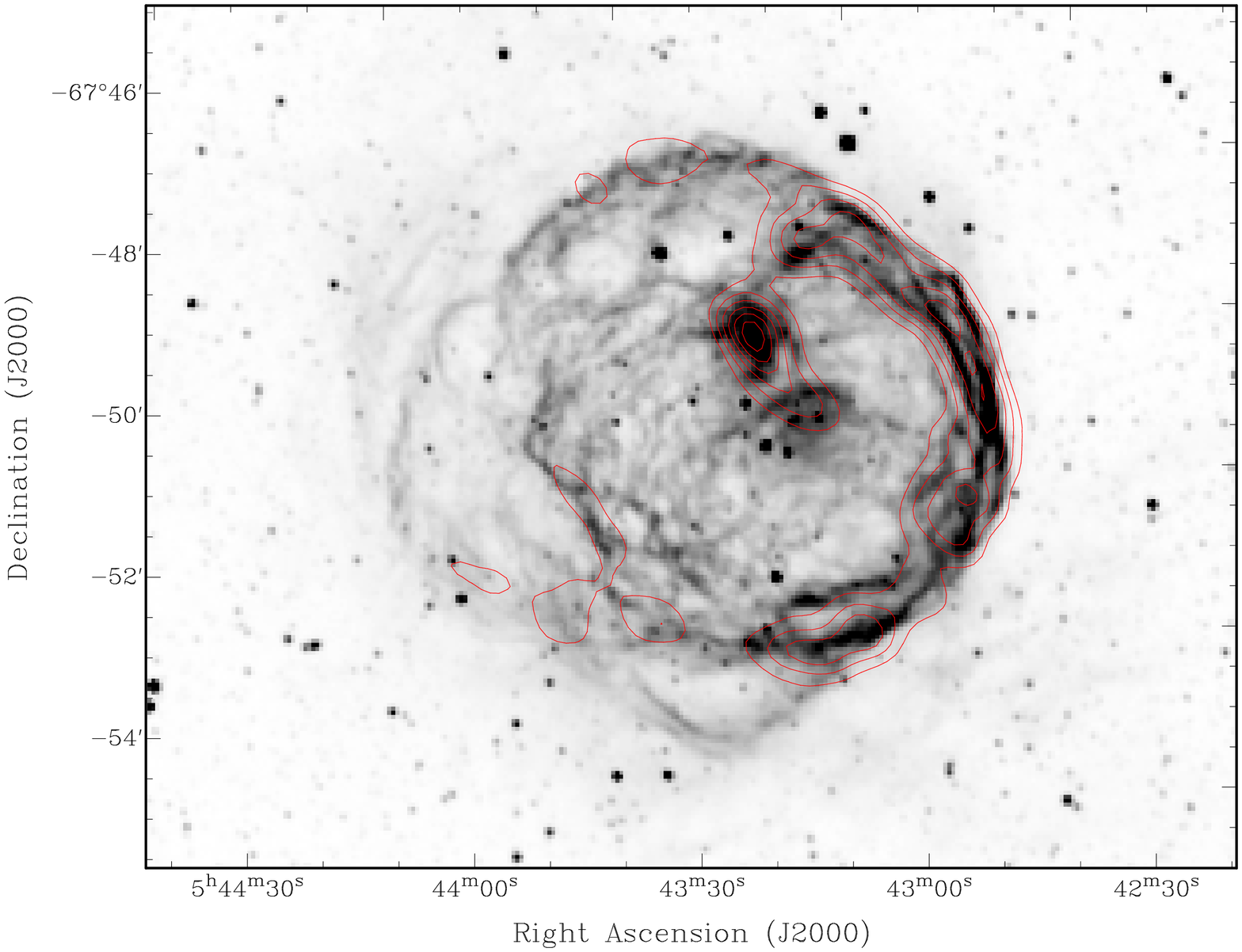}~\includegraphics[angle=-90,width=.33\textwidth]{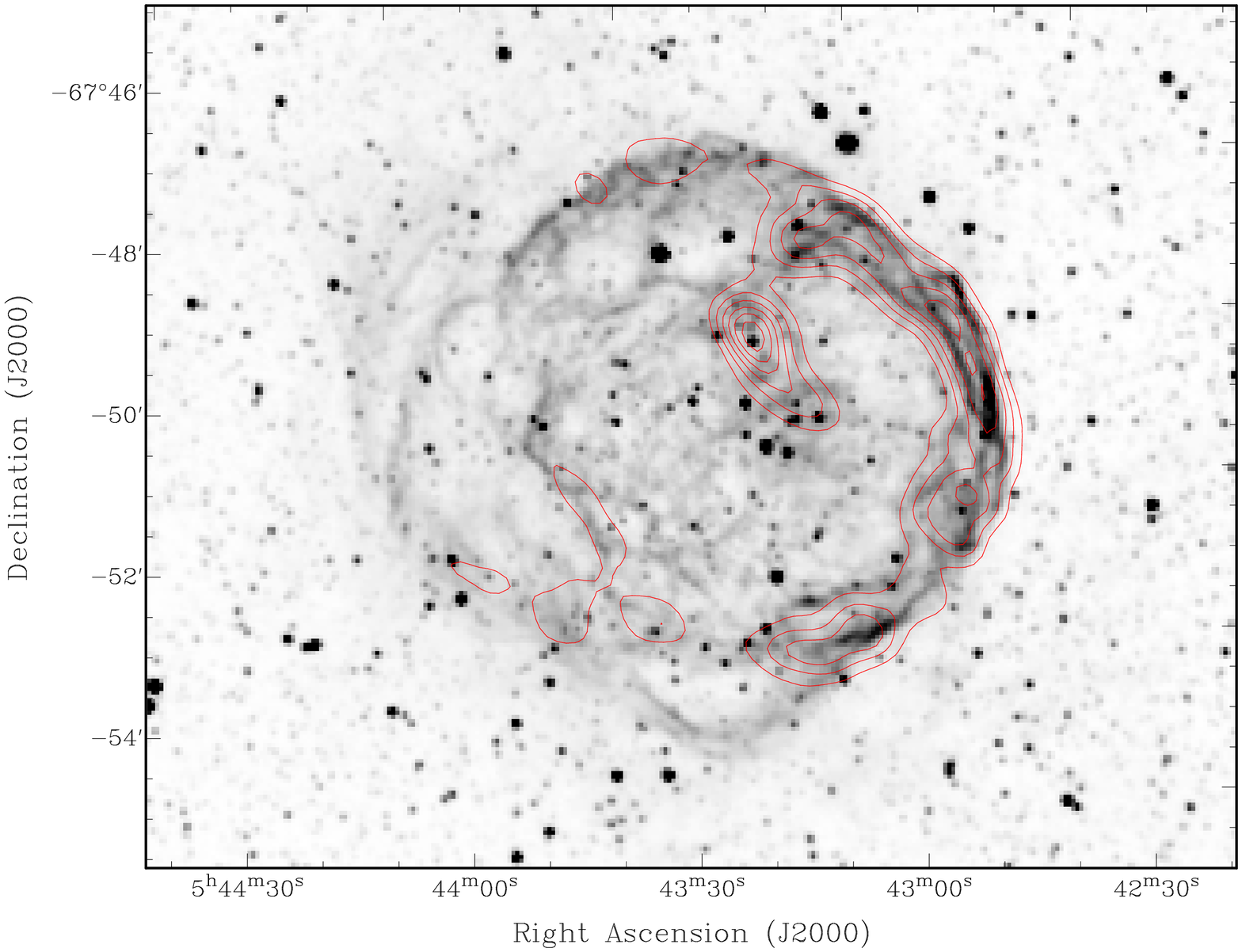}~\includegraphics[angle=-90,width=.33\textwidth]{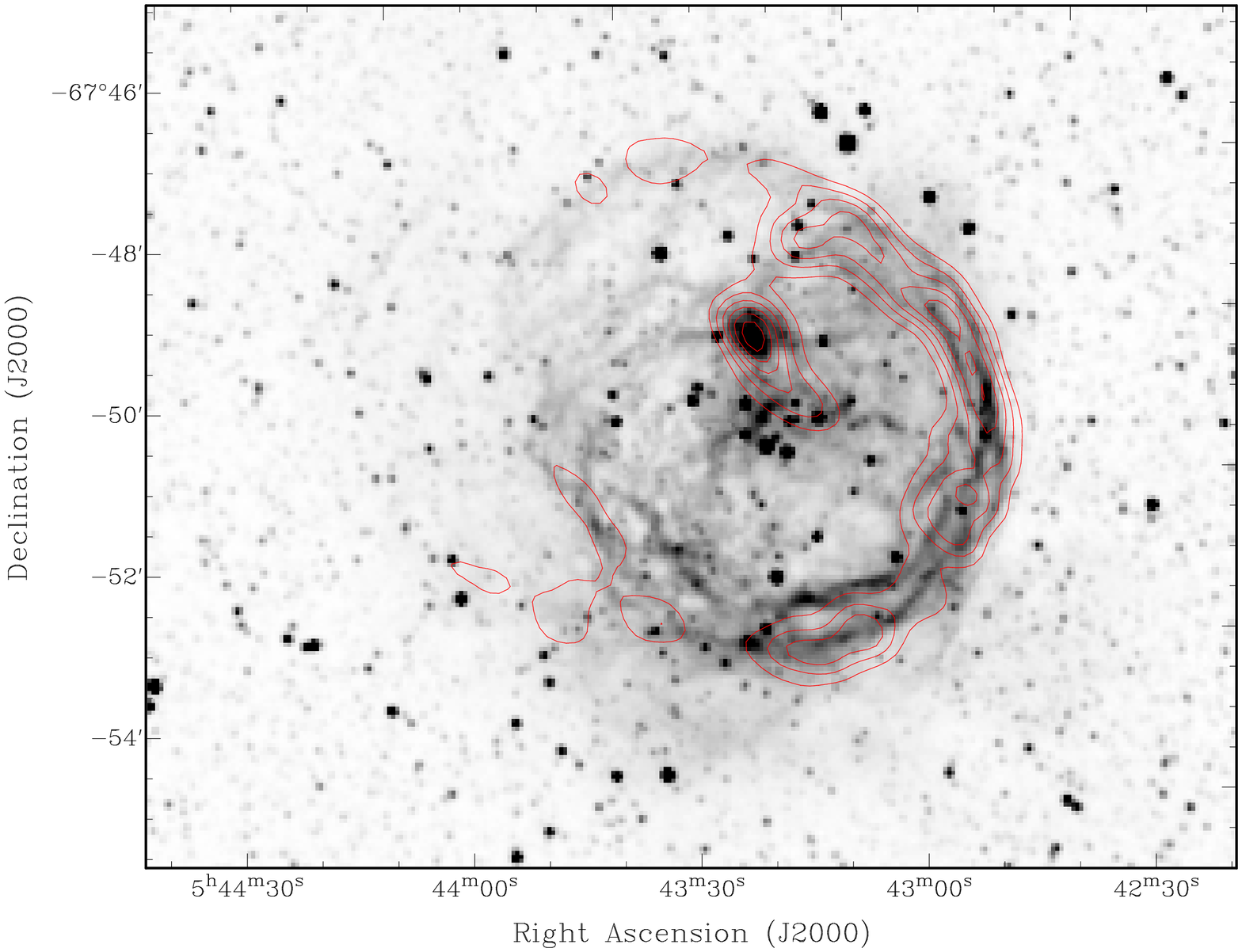}\vspace{-0.15 cm}}
 \caption{MCELS optical images (\Halpha\ (left), \SII\ (middle), \OIII\ (right)) of \SBN70\ overlaid with 6-cm contours at 3, 6, 9, 12, 15 and 18$\sigma$ ($\sigma$=0.12~mJy~beam$^{-1}$).}
 \label{fig:4}
\end{figure*}

The Magellanic Cloud Emission Line Survey (MCELS) is a survey of the  two  nearest  galaxies, the Small Magellanic Cloud and the LMC. The goal of the project is to trace the ionized gas in the Magellanic Clouds using narrow-band filters (Hydrogen (\Halpha\ 6563~\AA), Sulfur (\SII\ 6724~\AA) and Oxygen (\OIII\ 5007~\AA)). The survey is a joint project of the Cerro Tololo Inter-American Observatory (Chile) and the University of Michigan using the CTIO Curtis/Schmidt Telescope.  The detector used was a thinned, back-side illuminated Tek 2048$\times$2048 CCD with 24 micron pixels, giving a 1.35\degr\ field of view at a scale of 2.4\arcsec\,pixel$^{-1}$ with a resulting angular resolution of approximately 4.6\arcsec. Two slightly offset images were obtained through each filter to allow for cosmic ray rejection and bad pixel replacement. The total integration times were 1200~s in the \OIII\ and \SII\ images and 600~s in the \Halpha\ images per field. The data were reduced using a modified version of the Essence/SuperMaCHO pipeline for overscan correction, bias subtraction, and flat-field correction. Final astrometric solutions were derived using the IRAF software ``ccfind'' and ``ccmap'' using the UCAC catalog. At this time, the images were re-sampled to have 2$^{\prime\prime}$~$\times$~2$^{\prime\prime}$ pixels with an rms error of approximately 0.1 pixel (0.2$^{\prime\prime}$). The individual frames for each field were then aligned using the astrometric solutions and multiple exposures in each filter were combined using a median filter. Further details regarding the MCELS are given by Smith et al. (2006), \citet{2012ApJ...755...40P} and at http://www.ctio.noao.edu/mcels. Here, for the first time, we present optical images (Figures~\ref{fig:3} and~\ref{fig:4}) of this object in combination with our new radio-continuum and X-ray data.

\section{Results and Discussion}
 \label{s:resdis}

 \subsection{Radio-continuum}
\SBN70\ has a roughly circular morphology centered at \mbox{RA(J2000)=5\h43\m22\fs9}, \mbox{Dec(J2000)=--67\arcdeg50\arcmin57\arcsec} with a measured diameter at 6~cm of 104$\pm$1~pc. The diameter was determined using the \textsc{karma} tool {\sc kpvslice} to estimate the extension of \SBN70\ at 6~cm at the 3$\sigma$ noise level ($\sim$1~mJy). Overall, the optical and radio-continuum emissions follow each other.

\begin{figure*}
\centerline{\includegraphics[angle=90,width=1.0\columnwidth,clip=true, trim=1.5cm 4.2cm 3.25cm 3cm]{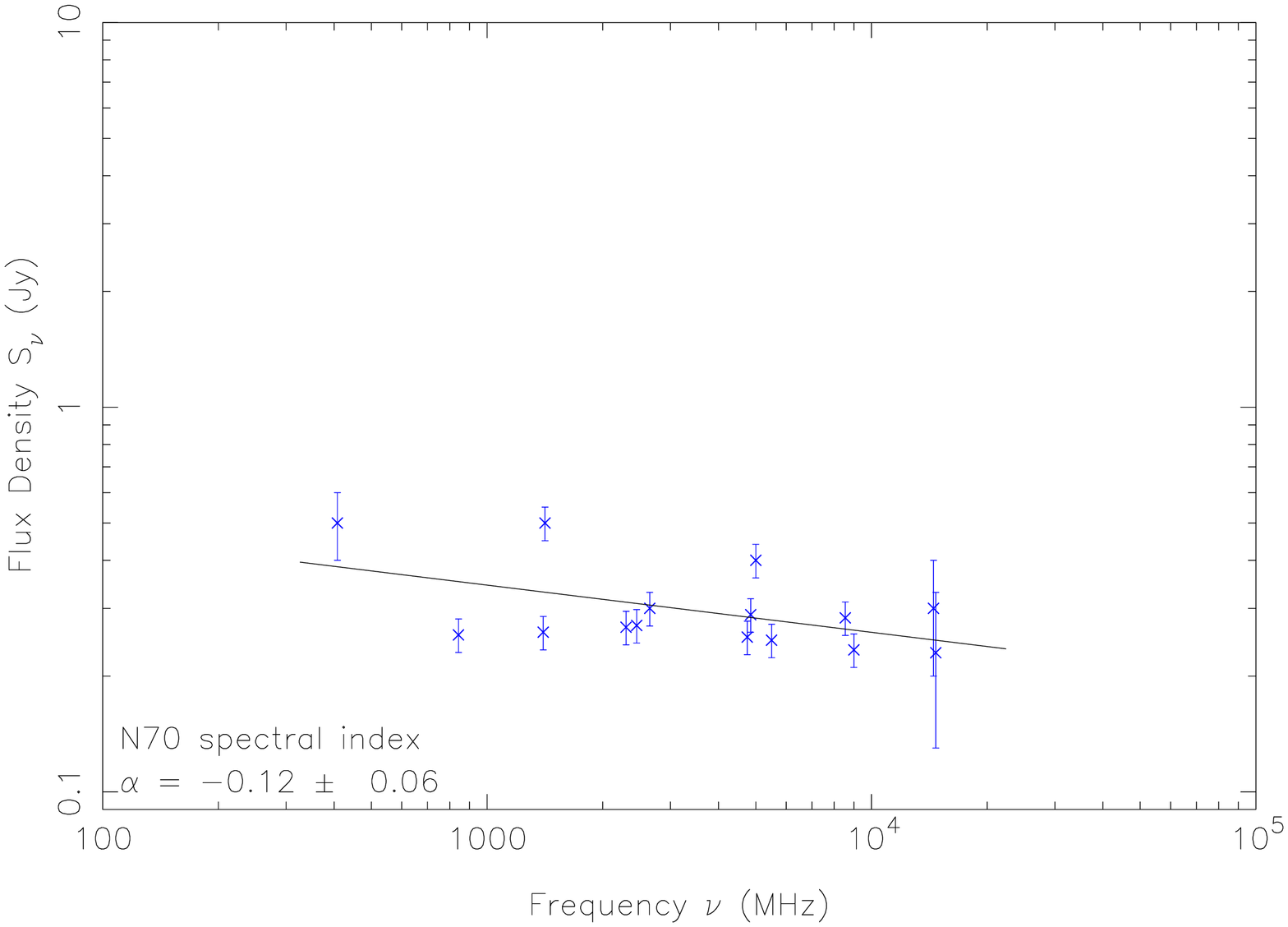}}
 \caption{Radio-continuum spectrum of \SBN70. The integrated flux densities at corresponding radio frequencies are listed in Table~\ref{tbl:3}.}
 \label{fig:5}
\end{figure*}

In order to estimate the radio spectral energy distribution for this object, we used our new integrated flux density measurements at various radio frequencies along with the 408~MHz measurement from \citet{1981ApJ...250..103D}, as well as various {\it Parkes} and other \atca\ estimates \citep{1995A&AS..111..311F,1996A&AS..120...77F,1998AAS..130..421F,2006MNRAS.370..363H,2007MNRAS.382..543H,2009MNRAS.399..769F}. We list these flux density measurements at various frequencies in Table~\ref{tbl:3} and then plot the \SBN70\ spectral index $\alpha$ in Figure~\ref{fig:5}. The overall radio-continuum spectral index of \SBN70\ is flat ($\alpha$=--0.12$\pm$0.06).

\begin{figure*}
 \includegraphics[angle=-90,width=1.0\columnwidth]{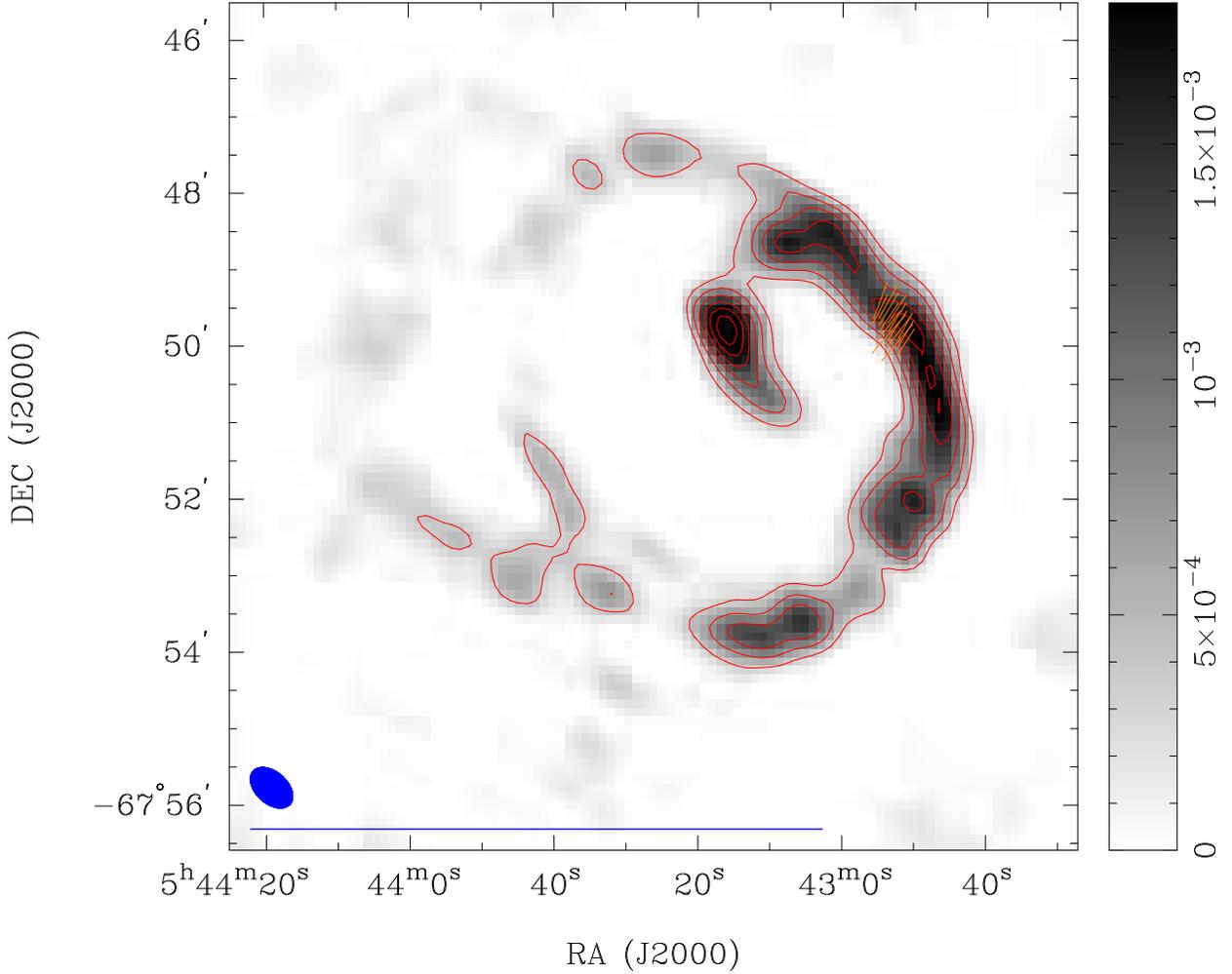}
 \caption{6-cm (5.5~GHz) ATCA observations of \SBN70. The blue ellipse in the lower-left corner represents the synthesised beam width of 39\,\arcsec$\times$25\arcsec at PA=48$^{\circ}$. The length of the vectors represents the fractional polarised intensity at each pixel position, and their orientations indicate the mean PA of the electric field (averaged over the observing bandwidth, not corrected for any Faraday rotation). The blue line below the beam ellipse represents the length of a polarisation vector of 100\%. The sidebar quantifies the pixel map and its units are Jy~beam$^{-1}$. The maximum fractional polarisation is 9\%$\pm1\%$. Contours at 3, 6, 9, 12, 15 and 18$\sigma$ ($\sigma$~=~0.12~mJy~beam$^{-1}$).}
 \label{fig:6}
\end{figure*}

The linear polarisation images for each frequency were created using \textit{Q} and \textit{U} parameters. The 6~cm image reveals some moderate linear polarisation with a mean fractional polarisation of 8.8\% $\pm1.1\%$ (no reliable detection could be made at 3~cm) in a small area of the north west limb (Figure~\ref{fig:6}). This linear polarisation and the moderate X-ray emission enhancement in the vicinity is indicative of an SN having occurred  close to the superbubble shell within $10^4$ years \citep{1990ApJ...365..510C}. This SN would certainly have provided chemical enrichment of the nebulous material. It would also contribute, in some way, to the overall energy that has shaped this area but the {exact} influence of the SN on current \SBN70\ structure is difficult to establish. \SBN70\ exhibits properties of an \HII\ region but the \SII\ to \Halpha\ ratio on the eastern side is higher than would be expected for a classic \HII\ region (Figures~\ref{fig:3} and \ref{fig:4}). The young stellar objects found in the cluster of stars within \SBN70\ are not massive enough to be wholly responsible for the bubble-like structure, even though they are certainly contributing to the energy within the structure (Zhang et al. 2014, submitted).

 \subsection{X-ray}

\begin{figure*}[t]
\begin{center}
\includegraphics[height=\hsize,angle=-90]{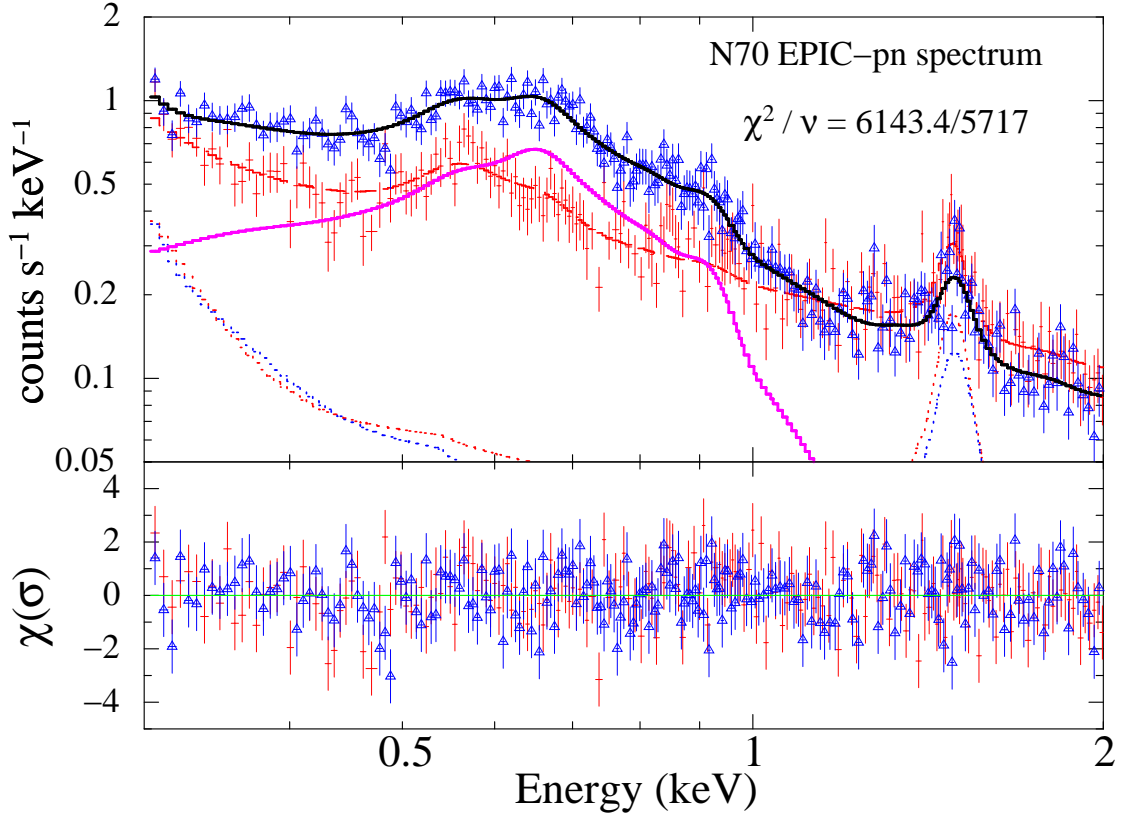}
\end{center}
\caption{EPIC-pn spectrum of \SBN70. Data extracted from the source region are shown by blue data points, with the total (source+background) model as the solid black line. The red and blue dash-dotted lines show the instrumental background model measured in the background and source extraction regions, respectively. The X-ray plus instrumental background data and model are shown by the red points and dashed line.  The thick magenta line shows the source emission component. The residuals are shown in terms of $\sigma$ in the bottom panel, where blue and red points are for the source and background spectra, respectively.}
\label{fig:7}
\end{figure*}

The majority of the X-ray emission appears to be in the softest (0.3--0.7~keV) energy band (Figure~\ref{fig:7}). To search for harder X-ray emission associated with \SBN70\ we created images in the 2--12~keV band, subtracted detector background and applied vignetting corrections as for the softer bands. In particular, subtraction of detector background is important as it dominates at high energies. We carefully inspected our images but find no evidence for diffuse emission above 2~keV as implied in \citet{2011RMxAC..40..199R}. The morphology in X-rays closely follows that in radio and optical, with most X-ray emission confined within the bright shell seen at longer wavelengths. A small deviation from circular morphology is observed towards the east of the SB, with fainter X-ray emission extending beyond the (broken) radio shell. There, it also correlates with fainter H$\alpha$ emission. We note that this is reminiscent of the  ``blister'' morphological classification examined in details by \citet{2012ApJ...755...40P}. They use ionisation-parameter mapping, looking at the \SII/\OIII\ ratio obtained from MCELS data to estimate the optical depth of \HII\ regions. They found a region optically thick on the east and thin on the west. This points to different conditions between the two directions. We also note that the impact of the hot gas leakage and blisters development on the shell dynamics has been recently discussed by \cite{2013MNRAS.435.3600P}.

\begin{figure*}
\centerline{\includegraphics[angle=270,width=1\textwidth,clip=true, trim=0.75cm 4.2cm 0.55cm 5.625cm]{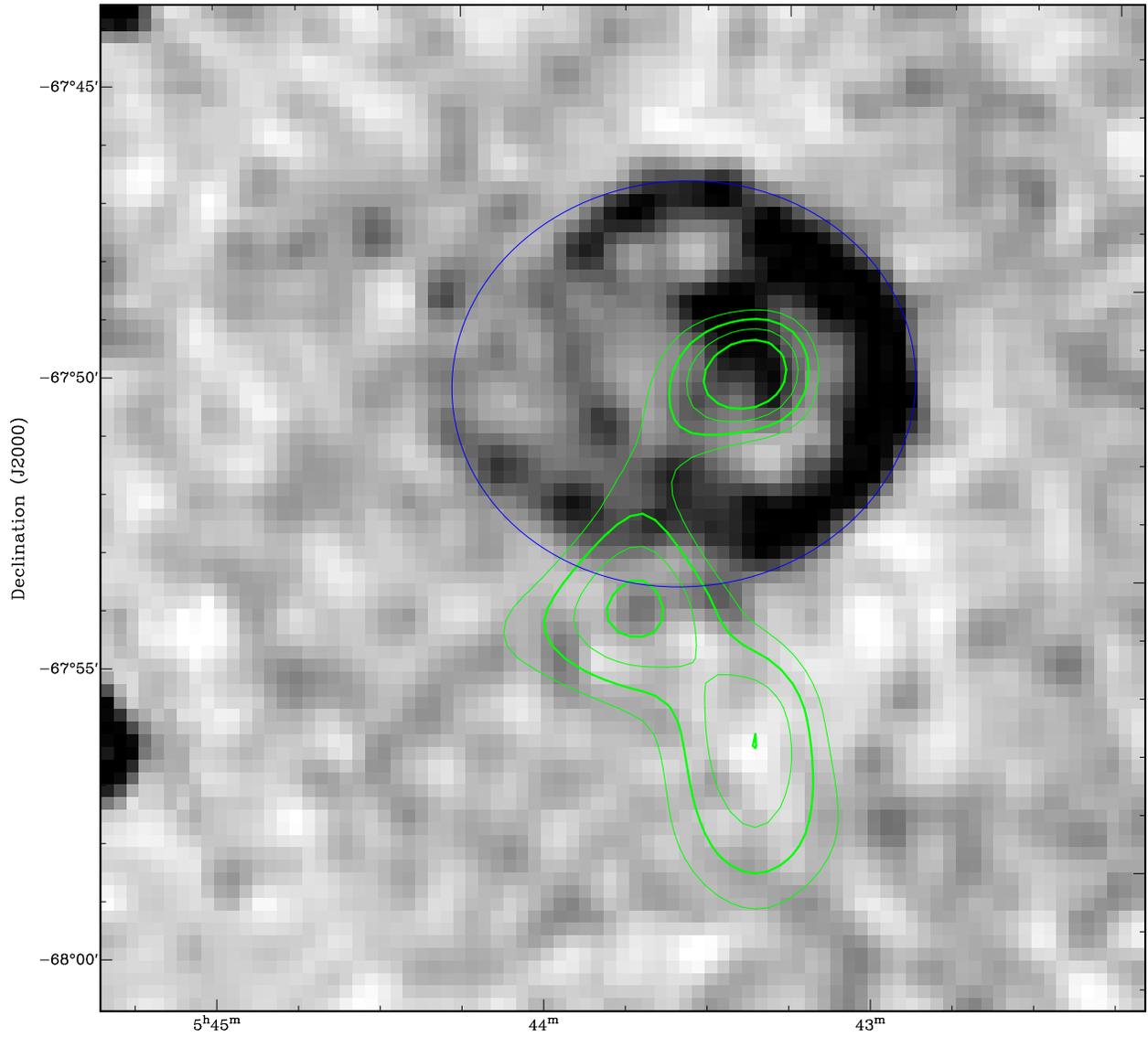}.}
\caption{\SBN70\ at 20~cm overlaid with the velocity-integrated intensity of the CO~(1Ð0) line obtained by \Nanten\ (green) at 0.75, 1, 1.25 and 1.5~K~km~s$^{-1}$ showing the association between the denser environment and the emission from the SB. The blue circle indicates the measured extent of the SB.}
\label{fig:8}
\end{figure*}

In addition, the X-ray surface brightness is slightly enhanced on the west, where the optical emission inside the shell is also increased. Radio emission at this position, correlating with the small molecular cloud seen in \Nanten\ CO~(1-0) emission \citep{2008ApJS..178...56F,2009ApJS..184....1K} (Figure~\ref{fig:8}), indicates interaction with higher densities, probably explaining the X-ray enhancement.

The X-ray spectrum of \SBN70\ is shown in Figure~\ref{fig:7}, on top of the modelled background. To fit the source spectrum we used a single-temperature model assuming collisional ionisation equilibrium (CIE). This utilises the Astrophysical Plasma Emission Code (APEC) in its most recent version available (v2.0.2), which includes updated atomic data \citep{2012ApJ...756..128F}. Possible effects of non-equilibrium ionisation (NEI) were also investigated by trying a plane-parallel shock model \citep[][\emph{vpshock} in XSPEC]{2001ApJ...548..820B}.

We accounted for Galactic foreground absorption towards \SBN70\ by a photoelectric absorption model (\emph{phabs} in XSPEC) at solar metallicity, with cross-sections taken from \citet{1992ApJ...400..699B}. The foreground column density was fixed at \mbox{$N_{H\ }= 6.3 \times\ 10^{20}$ cm$^{-2}$} \citep[based on the \HI\ measurement of][]{1990ARA&A..28..215D}. An additional absorption column to model absorption by atomic gas in the LMC was included, with a half-solar metallicity \citep{1992ApJ...384..508R}. Various abundance patterns were tried, such as scaling the solar values \citep[by number, from the table of][]{2000ApJ...542..914W} by a single fraction, fixing them at the abundances measured by \citet{1992ApJ...384..508R}, or leaving the abundances of oxygen and iron free to vary.

Purely thermal models adequately fit the soft spectrum, and the lack of harder emission (above 1~keV) is confirmed by the spectral analysis. We list the best-fit parameters in Table~\ref{tbl:4}. Plasma temperatures found are $kT \sim$ 0.25~keV. The temperature and $N_{H {\rm LMC}}$ are correlated: Equally acceptable fits are allowed both for a lower temperature (0.22~keV) with higher $N_{H {\rm LMC}}$ ($1.2 \times 10^{21}$~cm$^{-2}$) or for a higher temperature of 0.28~keV and a more moderate absorption ($0.3 \times 10^{21}$~cm$^{-2}$). We take that as an indication that the LMC-intrinsic absorption column is poorly constrained by the X-ray data alone. The ionisation age ($\tau = n_e t$) is large (more than 10$^{12}$~s\,cm$^{-3}$), suggesting that NIE effects are small or negligible. We find no evidence for abundances markedly different from LMC values, as the fits are consistent with a metallicity of 0.3--0.5. In particular, the only abundance ratio to which we have access in our data (O/Fe) is consistent, within the uncertainties, with that found in the LMC ISM. With the range of absorption column densities obtained from our spectral fitting, we measure an absorption-corrected X-ray luminosity (0.2--5~keV) between $2.3 \times 10^{35}$~erg\,s$^{-1}$ and $2.8 \times 10^{35}$~erg\,s$^{-1}$. As \citet{1990ApJ...365..510C} already found, this is one order of magnitude more than expected from a bubble model.

 \subsection{Analysis of the pressure budget of \SBN70}

At present, there are debates \citep[e.g.]{2013arXiv1309.5421L,2013ApJ...765...43S} on the impact that radiation pressure provides on the dynamics of the gas around young stellar clusters. Here, we estimate the pressure contributions by hot and warm gas and by dust-processed radiation.

The density of the X-ray phase can be derived from the spectral fits, using the Emission Measure $EM = n_{\rm X,e}~n_H~f~V$, with the electron density \mbox {$n_{\rm X,e}$ $\sim$1.2} times the hydrogen density $n_{\rm H}$. As for the volume $V$, we use a sphere of 50~pc radius. Given the morphology seen in X-ray it is likely that this volume is an upper limit (and consequently the derived density is a lower limit). From the result of the APEC model we have \mbox {$n_{\rm H}$ = 0.05 $f^{-1/2}$ cm$^{-3}$}. For the thermal pressure we take an ideal gas law \mbox {$P_{\rm X}$ = 1.9 $n_{\rm X}~k~T_{\rm X}$} \citep[Sect.~3.4]{2013arXiv1309.5421L} and then the estimated pressure is \mbox {$4.56 \times 10^{-11}~f^{-1/2}$~dyn~cm$^{-2}$ (where $f$ is he filling factor for which we assume a value of 1).} We compare this to the values of other \HII\ regions in \citet[Fig.~8 and Table~7]{2013arXiv1309.5421L} and find that it aligns well with the rest of the LMC sample.

We also estimate the warm gas pressure using \citep[Sect.~3.3]{2013arXiv1309.5421L}, as \mbox{$P_{\rm \HII} \sim 2 n_{\rm \HII,e}~k~T_{\rm \HII}$} with $T_{\rm \HII}$=10\,000~K. The electron density of the warm phase was estimated from spectroscopy \citep{1981A&A....97..342R} to be \mbox {$n_{\rm \HII,e}$=2.5~cm$^{-3}$}. We point out that $n_{\HII,e}$ is at least in the low density limit of $<$100~cm$^{-3}$. The $n_{\rm \HII,e}$ could be better determined using the 3-cm emission assuming it is dominated by free-free emission. However, we emphasise that at least a fraction of the flux is non-thermal and estimates could be misleading. Nevertheless, taking $n_{\rm \HII,e}$ of 2.5--100~cm$^{-3}$, we estimate that $P_{\rm \HII}$ ranges from 0.69 to 27.6$\times 10^{-11}$~dyn~cm$^{-2}$. Comparing with $P_{\rm X}$ of $4.56 \times 10^{-11}$~dyn~cm$^{-2}$, we find that the thermal pressure is slightly higher/lower than $P_{\rm \HII}$, as found in \citep{2013arXiv1309.5421L}. As $P_{\rm X}$/$P_{\rm \HII}$ is not $<<$1, we point that there is probably no significant hot gas leakage.

\citet{2014arXiv1402.2631S} estimated the infrared fluxes of N70. They found that the  interstellar radiation field (ISRF) of N70 has relatively
little variation with an average of U=6.825 (where the U is the dimensionless scale factor of energy density defined in  \citet[Sect.~3.2]{2013arXiv1309.5421L}.
This translates in a dust-processed radiation pressure of $P_{\rm IR} \sim 0.2 \times 10^{-11}$~dyn~cm$^{-2}$. This result is significantly smaller than any other \HII\ region analysed by \citet[Table~7; Col.~3]{2013arXiv1309.5421L} as the dust emission for \SBN70\ is likely driven by the shell and not the central region as suggested in \citet{2014arXiv1402.2631S}.

\citet[Sect.~3.1]{2013arXiv1309.5421L} implied that the direct radiation pressure ($P_{\rm dir}$) is generally small (more than an order of magnitude below $P_{\rm \HII}$ or $P_{\rm X}$). We note that the output of the most massive stars in \SBN70\ would not significantly differ from other \HII\ regions in the LMC apart from 30~Dor.

\section{Conclusion}
\label{con}

We present a multi-frequency study of the LMC SB \SBN70. With a diameter of 104~pc, \SBN70 is one of the most prominent objects in the entire LMC. The morphology of \SBN70\ in X-rays closely follows that in radio and optical, with most X-ray emission confined within the bright shell. The majority of the X-ray emission appears to be in the softest (0.3--0.7~keV) energy band. We do not detect any harder non-thermal X-ray emission.

We estimate the radio spectral index of \SBN70\ to be $\alpha$=--0.12$\pm$0.06 which is typical for non-SNR objects. However, we also detect limited polarisation with a maximum fractional polarisation of 9\% in a small area of the north west limb {indicative of an SN in the last $10^4$ years \citep{1990ApJ...365..510C}. We report an absorption-corrected X-ray luminosity (0.2--5~keV) between $2.3 \times 10^{35}$ erg\,s$^{-1}$ and $2.8 \times 10^{35}$ erg\,s$^{-1}$. Finally, we examine the pressure output and find that the hot ($P_{\rm X}$) and warm ($P_{\rm \HII}$) phase are consistent with other studied \HII\ regions while the dust-processed radiation pressure ($P_{\rm IR}$) is significantly smaller than in any other object studied in \citet{2013arXiv1309.5421L}.}

\acknowledgments
The ATCA is part of the Australia Telescope National Facility which is funded by the Commonwealth of Australia for operation as a National Facility managed by CSIRO. Based on observations obtained with \xmm, an ESA science mission with instruments and contributions directly funded by ESA Member States and NASA. The \xmm\ project is supported by the Bundesministerium f\"ur Wirtschaft und Technologie /Deutsches Zentrum f\"ur Luft- und Raumfahrt (BMWi/DLR, FKZ~50~OX~0001) and the Max-Planck Society. Cerro Tololo Inter-American Observatory (\CTIO) is operated by the Association of Universities for Research in Astronomy Inc. (AURA), under a cooperative agreement with the National Science Foundation (NSF) as part of the National Optical Astronomy Observatories (NOAO). We gratefully acknowledge the support of CTIO and all the assistance which has been provided in upgrading the Curtis Schmidt telescope. The MCELS is funded through the support of the Dean B. McLaughlin fund at the University of Michigan and through NSF grant 9540747. P.~M. acknowledge support from the BMWi/DLR grant FKZ~50 OR 1201. This research has made use of Aladin, SIMBAD and VizieR, operated at the CDS, Strasbourg, France. We used the {\sc karma} software package developed by the ATNF. {We thank the referee for numerous helpful comments that have greatly improved the quality of this paper.}


{\it Facilities:} \facility{\atca}, \facility{\xmm}, \facility{\CTIO}, \facility {\Nanten}.

\bibliographystyle{apj}
\bibliography{N70-REFs-FIN}

\begin{thebibliography}{63}
\expandafter\ifx\csname natexlab\endcsname\relax\def\natexlab#1{#1}\fi

\bibitem[{{Arnaud}(1996)}]{1996ASPC..101...17A}
{Arnaud}, K.~A. 1996, in Astronomical Society of the Pacific Conference Series,
  Vol. 101, Astronomical Data Analysis Software and Systems V, ed.
  {G.~H.~Jacoby \& J.~Barnes}, 17

\bibitem[{{Balucinska-Church} \& {McCammon}(1992)}]{1992ApJ...400..699B}
{Balucinska-Church}, M., \& {McCammon}, D. 1992, \apj, 400, 699

\bibitem[{{Boji{\v c}i{\'c}} {et~al.}(2007){Boji{\v c}i{\'c}}, {Filipovi{\'c}},
  {Parker}, {Payne}, {Jones}, {Reid}, {Kawamura}, \&
  {Fukui}}]{2007MNRAS.378.1237B}
{Boji{\v c}i{\'c}}, I.~S., {Filipovi{\'c}}, M.~D., {Parker}, Q.~A., {et~al.}
  2007, \mnras, 378, 1237

\bibitem[{{Borkowski} {et~al.}(2001){Borkowski}, {Lyerly}, \&
  {Reynolds}}]{2001ApJ...548..820B}
{Borkowski}, K.~J., {Lyerly}, W.~J., \& {Reynolds}, S.~P. 2001, \apj, 548, 820

\bibitem[{{Bozzetto} {et~al.}(2010){Bozzetto}, {Filipovic}, {Crawford},
  {Bojicic}, {Payne}, {Medik}, {Wardlaw}, \& {De~Horta}}]{2010SerAJ.181...43B}
{Bozzetto}, L.~M., {Filipovic}, M.~D., {Crawford}, E.~J., {et~al.} 2010,
  Serbian Astronomical Journal, 181, 43

\bibitem[{{Bozzetto} {et~al.}(2012{\natexlab{a}}){Bozzetto}, {Filipovic},
  {Crawford}, {De~Horta}, \& {Stupar}}]{2012SerAJ.184...69B}
{Bozzetto}, L.~M., {Filipovic}, M.~D., {Crawford}, E.~J., {De~Horta}, A.~Y., \&
  {Stupar}, M. 2012{\natexlab{a}}, Serbian Astronomical Journal, 184, 69

\bibitem[{{Bozzetto} {et~al.}(2012{\natexlab{b}}){Bozzetto}, {Filipovic},
  {Crawford}, {Payne}, {De~Horta}, \& {Stupar}}]{2012RMxAA..48...41B}
{Bozzetto}, L.~M., {Filipovic}, M.~D., {Crawford}, E.~J., {et~al.}
  2012{\natexlab{b}}, \rmxaa, 48, 41

\bibitem[{{Bozzetto} {et~al.}(2012{\natexlab{c}}){Bozzetto}, {Filipovi{\'c}},
  {Crawford}, {Haberl}, {Sasaki}, {Uro{\v s}evi{\'c}}, {Pietsch}, {Payne},
  {De~Horta}, {Stupar}, {Tothill}, {Dickel}, {Chu}, \&
  {Gruendl}}]{2012MNRAS.420.2588B}
{Bozzetto}, L.~M., {Filipovi{\'c}}, M.~D., {Crawford}, E.~J., {et~al.}
  2012{\natexlab{c}}, \mnras, 420, 2588

\bibitem[{{Bozzetto} {et~al.}(2013){Bozzetto}, {Filipovi{\'c}}, {Crawford},
  {Sasaki}, {Maggi}, {Haberl}, {Uro{\v s}evi{\'c}}, {Payne}, {De Horta},
  {Stupar}, {Gruendl}, \& {Dickel}}]{2013MNRAS.432.2177B}
---. 2013, \mnras, 432, 2177

\bibitem[{{\v{C}ajko} {et~al.}(2009){\v{C}ajko}, {Crawford}, \&
  {Filipovic}}]{2009SerAJ.179...55C}
{\v{C}ajko}, K.~O., {Crawford}, E.~J., \& {Filipovic}, M.~D. 2009, Serbian
  Astronomical Journal, 179, 55

\bibitem[{{Chu}(1997)}]{1997AJ....113.1815C}
{Chu}, Y.-H. 1997, \aj, 113, 1815

\bibitem[{{Chu} \& {Mac Low}(1990)}]{1990ApJ...365..510C}
{Chu}, Y.-H., \& {Mac Low}, M.-M. 1990, \apj, 365, 510

\bibitem[{{Crawford} {et~al.}(2008{\natexlab{a}}){Crawford}, {Filipovic},
  {De~Horta}, {Stootman}, \& {Payne}}]{2008SerAJ.177...61C}
{Crawford}, E.~J., {Filipovic}, M.~D., {De~Horta}, A.~Y., {Stootman}, F.~H., \&
  {Payne}, J.~L. 2008{\natexlab{a}}, Serbian Astronomical Journal, 177, 61

\bibitem[{{Crawford} {et~al.}(2010){Crawford}, {Filipovi{\'c}}, {Haberl},
  {Pietsch}, {Payne}, \& {De~Horta}}]{2010A&A...518A..35C}
{Crawford}, E.~J., {Filipovi{\'c}}, M.~D., {Haberl}, F., {et~al.} 2010, \aap,
  518, A35

\bibitem[{{Crawford} {et~al.}(2008{\natexlab{b}}){Crawford}, {Filipovic}, \&
  {Payne}}]{2008SerAJ.176...59C}
{Crawford}, E.~J., {Filipovic}, M.~D., \& {Payne}, J.~L. 2008{\natexlab{b}},
  Serbian Astronomical Journal, 176, 59

\bibitem[{{Danforth} \& {Blair}(2006)}]{2006ApJ...646..205D}
{Danforth}, C.~W., \& {Blair}, W.~P. 2006, \apj, 646, 205

\bibitem[{{Danziger} \& {Dennefeld}(1976)}]{1976PASP...88...44D}
{Danziger}, I.~J., \& {Dennefeld}, M. 1976, \pasp, 88, 44

\bibitem[{{Davies} {et~al.}(1976){Davies}, {Elliott}, \&
  {Meaburn}}]{1976MmRAS..81...89D}
{Davies}, R.~D., {Elliott}, K.~H., \& {Meaburn}, J. 1976, \memras, 81, 89

\bibitem[{{De~Horta} {et~al.}(2012){De~Horta}, {Filipovi{\'c}}, {Bozzetto},
  {Maggi}, {Haberl}, {Crawford}, {Sasaki}, {Uro{\v s}evi{\'c}}, {Pietsch},
  {Gruendl}, {Dickel}, {Tothill}, {Chu}, {Payne}, \&
  {Collier}}]{2012A&A...540A..25D}
{De~Horta}, A.~Y., {Filipovi{\'c}}, M.~D., {Bozzetto}, L.~M., {et~al.} 2012,
  \aap, 540, A25

\bibitem[{{Dickey} \& {Lockman}(1990)}]{1990ARA&A..28..215D}
{Dickey}, J.~M., \& {Lockman}, F.~J. 1990, \araa, 28, 215

\bibitem[{{Dopita} {et~al.}(1981){Dopita}, {Ford}, {McGregor}, {Mathewson}, \&
  {Wilson}}]{1981ApJ...250..103D}
{Dopita}, M.~A., {Ford}, V.~L., {McGregor}, P.~J., {Mathewson}, D.~S., \&
  {Wilson}, I.~R. 1981, \apj, 250, 103

\bibitem[{{Filipovi\'c} {et~al.}(1998{\natexlab{a}}){Filipovi\'c}, {Haynes},
  {White}, \& {Jones}}]{1998AAS..130..421F}
{Filipovi\'c}, M.~D., {Haynes}, R.~F., {White}, G.~L., \& {Jones}, P.~A.
  1998{\natexlab{a}}, \aaps, 130, 421

\bibitem[{{Filipovi\'c} {et~al.}(1995){Filipovi\'c}, {Haynes}, {White},
  {Jones}, {Klein}, \& {Wielebinski}}]{1995A&AS..111..311F}
{Filipovi\'c}, M.~D., {Haynes}, R.~F., {White}, G.~L., {et~al.} 1995, \aaps,
  111, 311

\bibitem[{{Filipovi\'c} {et~al.}(1998{\natexlab{b}}){Filipovi\'c}, {Jones},
  {White}, \& {Haynes}}]{1998PASA...15..128F}
{Filipovi\'c}, M.~D., {Jones}, P.~A., {White}, G.~L., \& {Haynes}, R.~F.
  1998{\natexlab{b}}, \pasa, 15, 128

\bibitem[{{Filipovi\'c} {et~al.}(1996){Filipovi\'c}, {White}, {Haynes},
  {Jones}, {Meinert}, {Wielebinski}, \& {Klein}}]{1996A&AS..120...77F}
{Filipovi\'c}, M.~D., {White}, G.~L., {Haynes}, R.~F., {et~al.} 1996, \aaps,
  120, 77

\bibitem[{{Filipovi{\'c}} {et~al.}(2009){Filipovi{\'c}}, {Cohen}, {Reid},
  {Payne}, {Parker}, {Crawford}, {Boji{\v c}i{\'c}}, {de Horta}, {Hughes},
  {Dickel}, \& {Stootman}}]{2009MNRAS.399..769F}
{Filipovi{\'c}}, M.~D., {Cohen}, M., {Reid}, W.~A., {et~al.} 2009, \mnras, 399,
  769

\bibitem[{{Foster} {et~al.}(2012){Foster}, {Ji}, {Smith}, \&
  {Brickhouse}}]{2012ApJ...756..128F}
{Foster}, A.~R., {Ji}, L., {Smith}, R.~K., \& {Brickhouse}, N.~S. 2012, \apj,
  756, 128

\bibitem[{{Fukui} {et~al.}(2008){Fukui}, {Kawamura}, {Minamidani}, {Mizuno},
  {Kanai}, {Mizuno}, {Onishi}, {Yonekura}, {Mizuno}, {Ogawa}, \&
  {Rubio}}]{2008ApJS..178...56F}
{Fukui}, Y., {Kawamura}, A., {Minamidani}, T., {et~al.} 2008, \apjs, 178, 56

\bibitem[{{Georgelin} {et~al.}(1983){Georgelin}, {Georgelin}, {Laval},
  {Monnet}, \& {Rosado}}]{1983A&AS...54..459G}
{Georgelin}, Y.~M., {Georgelin}, Y.~P., {Laval}, A., {Monnet}, G., \& {Rosado},
  M. 1983, \aaps, 54, 459

\bibitem[{{Gooch}(1996)}]{1996ASPC..101...80G}
{Gooch}, R. 1996, in Astronomical Society of the Pacific Conference Series,
  Vol. 101, Astronomical Data Analysis Software and Systems V, ed. G.~H.
  {Jacoby} \& J.~{Barnes}, 80

\bibitem[{{Grondin} {et~al.}(2012){Grondin}, {Sasaki}, {Haberl}, {Pietsch},
  {Crawford}, {Filipovi{\'c}}, {Bozzetto}, {Points}, \&
  {Smith}}]{2012A&A...539A..15G}
{Grondin}, M.-H., {Sasaki}, M., {Haberl}, F., {et~al.} 2012, \aap, 539, A15

\bibitem[{{Henize}(1956)}]{1956ApJS....2..315H}
{Henize}, K.~G. 1956, \apjs, 2, 315

\bibitem[{{Hughes} {et~al.}(2007){Hughes}, {Staveley-Smith}, {Kim}, {Wolleben},
  \& {Filipovi{\'c}}}]{2007MNRAS.382..543H}
{Hughes}, A., {Staveley-Smith}, L., {Kim}, S., {Wolleben}, M., \&
  {Filipovi{\'c}}, M. 2007, \mnras, 382, 543

\bibitem[{{Hughes} {et~al.}(2006){Hughes}, {Wong}, {Ekers}, {Staveley-Smith},
  {Filipovic}, {Maddison}, {Fukui}, \& {Mizuno}}]{2006MNRAS.370..363H}
{Hughes}, A., {Wong}, T., {Ekers}, R., {et~al.} 2006, \mnras, 370, 363

\bibitem[{{Inglis} \& {Kitchin}(1990)}]{1990MNRAS.246..358I}
{Inglis}, M.~D., \& {Kitchin}, C.~R. 1990, \mnras, 246, 358

\bibitem[{{Kawamura} {et~al.}(2009){Kawamura}, {Mizuno}, {Minamidani},
  {Filipovi{\'c}}, {Staveley-Smith}, {Kim}, {Mizuno}, {Onishi}, {Mizuno}, \&
  {Fukui}}]{2009ApJS..184....1K}
{Kawamura}, A., {Mizuno}, Y., {Minamidani}, T., {et~al.} 2009, \apjs, 184, 1

\bibitem[{{Lopez} {et~al.}(2013){Lopez}, {Krumholz}, {Bolatto}, {Prochaska},
  {Ramirez-Ruiz}, \& {Castro}}]{2013arXiv1309.5421L}
{Lopez}, L.~A., {Krumholz}, M.~R., {Bolatto}, A.~D., {et~al.} 2013, ArXiv
  e-prints

\bibitem[{{Lucke} \& {Hodge}(1970)}]{1970AJ.....75..171L}
{Lucke}, P.~B., \& {Hodge}, P.~W. 1970, \aj, 75, 171

\bibitem[{{Maggi} {et~al.}(2012){Maggi}, {Haberl}, {Bozzetto}, {Filipovi{\'c}},
  {Points}, {Chu}, {Sasaki}, {Pietsch}, {Gruendl}, {Dickel}, {Smith}, {Sturm},
  {Crawford}, \& {De~Horta}}]{2012A&A...546A.109M}
{Maggi}, P., {Haberl}, F., {Bozzetto}, L.~M., {et~al.} 2012, \aap, 546, A109

\bibitem[{{Mathis} {et~al.}(1983){Mathis}, {Mezger}, \&
  {Panagia}}]{1983A&A...128..212M}
{Mathis}, J.~S., {Mezger}, P.~G., \& {Panagia}, N. 1983, \aap, 128, 212

\bibitem[{{Mills} {et~al.}(1984){Mills}, {Turtle}, {Little}, \&
  {Durdin}}]{1984AuJPh..37..321M}
{Mills}, B.~Y., {Turtle}, A.~J., {Little}, A.~G., \& {Durdin}, J.~M. 1984,
  Australian Journal of Physics, 37, 321

\bibitem[{{Milne} {et~al.}(1980){Milne}, {Caswell}, \&
  {Haynes}}]{1980MNRAS.191..469M}
{Milne}, D.~K., {Caswell}, J.~L., \& {Haynes}, R.~F. 1980, \mnras, 191, 469

\bibitem[{{Oey}(1996{\natexlab{a}})}]{1996ApJ...467..666O}
{Oey}, M.~S. 1996{\natexlab{a}}, \apj, 467, 666

\bibitem[{{Oey}(1996{\natexlab{b}})}]{1996ApJS..104...71O}
---. 1996{\natexlab{b}}, \apjs, 104, 71

\bibitem[{{Oey} \& {Garc{\'{\i}}a-Segura}(2004)}]{2004ApJ...613..302O}
{Oey}, M.~S., \& {Garc{\'{\i}}a-Segura}, G. 2004, \apj, 613, 302

\bibitem[{{Oey} \& {Kennicutt}(1997)}]{1997MNRAS.291..827O}
{Oey}, M.~S., \& {Kennicutt}, Jr., R.~C. 1997, \mnras, 291, 827

\bibitem[{{Pellegrini} {et~al.}(2012){Pellegrini}, {Oey}, {Winkler}, {Points},
  {Smith}, {Jaskot}, \& {Zastrow}}]{2012ApJ...755...40P}
{Pellegrini}, E.~W., {Oey}, M.~S., {Winkler}, P.~F., {et~al.} 2012, \apj, 755,
  40

\bibitem[{{Pietrzy{\'n}ski} {et~al.}(2013){Pietrzy{\'n}ski}, {Graczyk},
  {Gieren}, {Thompson}, {Pilecki}, {Udalski}, {Soszy{\'n}ski}, {Koz{\l}owski},
  {Konorski}, {Suchomska}, {Bono}, {Moroni}, {Villanova}, {Nardetto},
  {Bresolin}, {Kudritzki}, {Storm}, {Gallenne}, {Smolec}, {Minniti}, {Kubiak},
  {Szyma{\'n}ski}, {Poleski}, {Wyrzykowski}, {Ulaczyk}, {Pietrukowicz},
  {G{\'o}rski}, \& {Karczmarek}}]{2013Natur.495...76P}
{Pietrzy{\'n}ski}, G., {Graczyk}, D., {Gieren}, W., {et~al.} 2013, \nat, 495,
  76

\bibitem[{{Pittard}(2013)}]{2013MNRAS.435.3600P}
{Pittard}, J.~M. 2013, \mnras, 435, 3600

\bibitem[{{Reyes-Iturbide} {et~al.}(2011){Reyes-Iturbide}, {Rosado},
  {Rodr{\'{\i}}guez-Gonz{\'a}lez}, {Vel{\'a}zquez}, \&
  {Ambrocio-Cruz}}]{2011RMxAC..40..199R}
{Reyes-Iturbide}, J., {Rosado}, M., {Rodr{\'{\i}}guez-Gonz{\'a}lez}, A.,
  {Vel{\'a}zquez}, P.~F., \& {Ambrocio-Cruz}, P. 2011, in Revista Mexicana de
  Astronomia y Astrofisica Conference Series, Vol.~40, Revista Mexicana de
  Astronomia y Astrofisica Conference Series, 199--199

\bibitem[{{Rodr{\'{\i}}guez-Gonz{\'a}lez}
  {et~al.}(2011){Rodr{\'{\i}}guez-Gonz{\'a}lez}, {Vel{\'a}zquez}, {Rosado},
  {Esquivel}, {Reyes-Iturbide}, \& {Toledo-Roy}}]{2011ApJ...733...34R}
{Rodr{\'{\i}}guez-Gonz{\'a}lez}, A., {Vel{\'a}zquez}, P.~F., {Rosado}, M.,
  {et~al.} 2011, \apj, 733, 34

\bibitem[{{Rosado} {et~al.}(1981){Rosado}, {Georgelin}, {Georgelin}, {Laval},
  \& {Monnet}}]{1981A&A....97..342R}
{Rosado}, M., {Georgelin}, Y.~P., {Georgelin}, Y.~M., {Laval}, A., \& {Monnet},
  G. 1981, \aap, 97, 342

\bibitem[{{Russell} \& {Dopita}(1992)}]{1992ApJ...384..508R}
{Russell}, S.~C., \& {Dopita}, M.~A. 1992, \apj, 384, 508

\bibitem[{{Sault} \& {Killeen}(2006)}]{2006Miriad}
{Sault}, B., \& {Killeen}, N. 2006, Miriad Users Guide (Australia Telescope
  National Facility)

\bibitem[{{Sault} \& {Wieringa}(1994)}]{1994A&AS..108..585S}
{Sault}, R.~J., \& {Wieringa}, M.~H. 1994, \aaps, 108, 585

\bibitem[{{Silich} \& {Tenorio-Tagle}(2013)}]{2013ApJ...765...43S}
{Silich}, S., \& {Tenorio-Tagle}, G. 2013, \apj, 765, 43

\bibitem[{{Slater} {et~al.}(2011){Slater}, {Oey}, {Li}, {Bernard},
  {Churchwell}, {Gordon}, {Indebetouw}, {Lawton}, {Meixner}, {Paradis}, \&
  {Reach}}]{2011ApJ...732...98S}
{Slater}, C.~T., {Oey}, M.~S., {Li}, A., {et~al.} 2011, \apj, 732, 98

\bibitem[{{Stephens} {et~al.}(2014){Stephens}, {Evans}, {Xue}, {Chu},
  {Gruendl}, \& {Segura-Cox}}]{2014arXiv1402.2631S}
{Stephens}, I.~W., {Evans}, J.~M., {Xue}, R., {et~al.} 2014, ArXiv e-prints

\bibitem[{{Str{\"u}der} {et~al.}(2001){Str{\"u}der}, {Briel}, {Dennerl},
  {Hartmann}, {Kendziorra}, {Meidinger}, {Pfeffermann}, {Reppin}, {Aschenbach},
  {Bornemann}, {Br{\"a}uninger}, {Burkert}, {Elender}, {Freyberg}, {Haberl},
  {Hartner}, {Heuschmann}, {Hippmann}, {Kastelic}, {Kemmer}, {Kettenring},
  {Kink}, {Krause}, {M{\"u}ller}, {Oppitz}, {Pietsch}, {Popp}, {Predehl},
  {Read}, {Stephan}, {St{\"o}tter}, {Tr{\"u}mper}, {Holl}, {Kemmer}, {Soltau},
  {St{\"o}tter}, {Weber}, {Weichert}, {von Zanthier}, {Carathanassis}, {Lutz},
  {Richter}, {Solc}, {B{\"o}ttcher}, {Kuster}, {Staubert}, {Abbey}, {Holland},
  {Turner}, {Balasini}, {Bignami}, {La Palombara}, {Villa}, {Buttler},
  {Gianini}, {Lain{\'e}}, {Lumb}, \& {Dhez}}]{2001A&A...365L..18S}
{Str{\"u}der}, L., {Briel}, U., {Dennerl}, K., {et~al.} 2001, \aap, 365, L18

\bibitem[{{Turner} {et~al.}(2001){Turner}, {Abbey}, {Arnaud}, {Balasini},
  {Barbera}, {Belsole}, {Bennie}, {Bernard}, {Bignami}, {Boer}, {Briel},
  {Butler}, {Cara}, {Chabaud}, {Cole}, {Collura}, {Conte}, {Cros}, {Denby},
  {Dhez}, {Di Coco}, {Dowson}, {Ferrando}, {Ghizzardi}, {Gianotti}, {Goodall},
  {Gretton}, {Griffiths}, {Hainaut}, {Hochedez}, {Holland}, {Jourdain},
  {Kendziorra}, {Lagostina}, {Laine}, {La Palombara}, {Lortholary}, {Lumb},
  {Marty}, {Molendi}, {Pigot}, {Poindron}, {Pounds}, {Reeves}, {Reppin},
  {Rothenflug}, {Salvetat}, {Sauvageot}, {Schmitt}, {Sembay}, {Short},
  {Spragg}, {Stephen}, {Str{\"u}der}, {Tiengo}, {Trifoglio}, {Tr{\"u}mper},
  {Vercellone}, {Vigroux}, {Villa}, {Ward}, {Whitehead}, \&
  {Zonca}}]{2001A&A...365L..27T}
{Turner}, M.~J.~L., {Abbey}, A., {Arnaud}, M., {et~al.} 2001, \aap, 365, L27

\bibitem[{{Wang} \& {Helfand}(1991)}]{1991ApJ...373..497W}
{Wang}, Q., \& {Helfand}, D.~J. 1991, \apj, 373, 497

\bibitem[{{Weaver} {et~al.}(1977){Weaver}, {McCray}, {Castor}, {Shapiro}, \&
  {Moore}}]{1977ApJ...218..377W}
{Weaver}, R., {McCray}, R., {Castor}, J., {Shapiro}, P., \& {Moore}, R. 1977,
  \apj, 218, 377

\bibitem[{{Wilms} {et~al.}(2000){Wilms}, {Allen}, \&
  {McCray}}]{2000ApJ...542..914W}
{Wilms}, J., {Allen}, A., \& {McCray}, R. 2000, \apj, 542, 914

\end{thebibliography}


\clearpage

\begin{table}
 \small
 \caption{Summary of Australia Telescope Online Archive (ATOA) observations from project C634 used in imaging \SBN70.} 
 \begin{tabular}{@{}l r c c c r r r@{}}
\tableline  
\multicolumn{1}{c}{Date} & \multicolumn{1}{c}{Scan time} &  \multicolumn{1}{c}{RA} & \multicolumn{1}{c}{Dec} & \multicolumn{1}{c}{Array} & \multicolumn{1}{c}{Frequencies} & \multicolumn{1}{c}{Bandwidth} & \multicolumn{1}{c}{Number of}\\
 & \multicolumn{1}{c}{(minutes)} & \multicolumn{1}{c}{(J2000)}& \multicolumn{1}{c}{(J2000)}& & \multicolumn{1}{c}{(MHz)} & \multicolumn{1}{c}{(MHz)} & \multicolumn{1}{c}{Channels}\\
\tableline
2011-Nov-16 & 44.6\p0\p0\p0 & 5\h 43\m 25\fs00 & --67\arcdeg50\arcmin57\farcs20 & EW367&5,500, 9,000 & 2048.0\p0\p0     & 2049\p0\p0\p0\\
2011-Nov-15 & 24.5\p0\p0\p0 & 5\h 43\m 25\fs00 & --67\arcdeg50\arcmin57\farcs20 & EW367&5,500, 9,000 & 2048.0\p0\p0     & 2049\p0\p0\p0\\
1997-Oct-06  & 10.0\p0\p0\p0 & 5\h 43\m 14\fs46 & --67\arcdeg51\arcmin23\farcs75 & 375&4,790, 8,640       & \p0128.0\p0\p0 & 33\p0\p0\p0\\
1997-Oct-05  & 10.0\p0\p0\p0 & 5\h 43\m 14\fs46 & --67\arcdeg51\arcmin23\farcs75 & 375&4,790, 8,640       & \p0128.0\p0\p0 & 33\p0\p0\p0\\
1997-Oct-02  & 80.0\p0\p0\p0 & 5\h 43\m 14\fs46 & --67\arcdeg51\arcmin23\farcs75 & 375&4,800, 8,640       & \p0128.0\p0\p0 & 33\p0\p0\p0\\
\tableline 
 \end{tabular}
 \label{tbl:1}
\end{table}

\begin{table}
 \small
 \caption{Summary of ATOA observations from project C587 used in imaging N\,70. All projects were observed at a frequency of 1380~MHz and a bandwidth of 128~MHz across 33 channels. All observations were centred at RA (J2000)=5\h43\m39\fs10 Dec (J2000)=--67\arcdeg50\arcmin6\farcs80.}
 \label{tbl:2} 
 \begin{tabular}{@{}c c c @{}}
\tableline
 Date & Scan time &  Array\\
 & (minutes) &  \\
\tableline
1997-Sep-04 & 138.3       &   1.5C \\
1997-Sep-03 & \p034.0    &   1.5C  \\
1997-Mar-31 & \p020.8    &   1.5D  \\
1997-Mar-27 & \p037.0    &   1.5D  \\
1997-Mar-26 & 175.0       &   1.5D  \\
1997-Jan-08 & \p0\p02.3 &   750D  \\
1997-Jan-07 & 126.9       &   750D  \\
1997-Jan-06 & \p098.2    &   750D  \\
1997-Jan-05 & 143.8       &   750D  \\
1996-Nov-28 & 129.3      &   750A  \\
1996-Nov-27 & 112.3      &   750A  \\
\tableline
 \end{tabular}
\end{table}

\begin{table*}[tb3]
 \small
 \caption{Measured integrated flux density of \SBN70.}
 \label{tbl:3} 
 \begin{tabular}{cclccl}
\hline
$\nu$       & $\lambda$ & Telescope & Beam Size & S$_\mathrm{Total}$  & Reference \\
(MHz)       & (cm)      &           & (\arcsec) & (mJy)               &           \\
\hline
\p0\p0408 & 73    & Molonglo    & 174$\times$204 & 500  & \cite{1981ApJ...250..103D}\\
\p0\p0843 & 36    & MOST        & 43$\times$43   & 256  & This work \\
\p01400   & 21    & ATCA        & 40$\times$40   & 260  & This work \\
\p01415   & 21    & Fleurs S.T. & 174$\times$204 & 500  & \cite{1981ApJ...250..103D}\\
\p02300   & 13    & Parkes      & 540$\times$540 & 268  & \cite{1998AAS..130..421F}\\
\p02450   & 12    & Parkes      & 540$\times$540 & 271  & \cite{1998AAS..130..421F}\\
\p02650   & 11    & Parkes      & 450$\times$450 & 300  & \cite{1981ApJ...250..103D}\\
\p04750   & \p06  & Parkes      & 294$\times$294 & 253  & \cite{1998AAS..130..421F}\\
\p04850   & \p06  & Parkes/PMN  & 294$\times$294 & 289  & \cite{1998AAS..130..421F}\\
\p05000   & \p06  & Parkes      & 264$\times$264 & 400  & \cite{1981ApJ...250..103D}\\
\p05500   & \p06  & ATCA        & 2.7$\times$2.1 & 248  & This work\\
\p08550   & \p03  & Parkes      & 162$\times$162 & 283  & \cite{1998AAS..130..421F}\\
\p09000   & \p03  & ATCA        & 2.7$\times$2.1 & 234  & This work \\
14500     & \p02  & Parkes      & 132$\times$132 & 300  & \cite{1981ApJ...250..103D}\\
 \hline
 \end{tabular}
\end{table*}

\begin{table*}[t]
 \small
 \caption{X-ray spectral results.}
 \label{tbl:4}
 \centering
 \begin{tabular}{l c c c c c c c c}
\tableline
\noalign{\smallskip}
Model & N$_{H\mathrm{\ LMC}}$ & $kT$ & $\tau$ & EM\tablenotemark{\dag} & O/H & Fe/H & $\chi^2 $ / dof & $L_X$\tablenotemark{\ddag}\\ &($10^{20}$ cm$^{-2}$)  & (eV) & ($10^{12}$ s\,cm$^{-3}$) &($10^{58}$ cm$^{-3}$)&
 & & & ($10^{35}$ erg\,s$^{-1}$) \\
\noalign{\smallskip}
\tableline
\noalign{\smallskip}
\textit{apec}  & 1.4 $(< 5.4)$ & 255$\pm 9$ & --- & 3.92$_{-1.58} ^{+2.58}$ &
\multicolumn{2}{c}{$Z = 0.27_{-0.08} ^{+0.10} Z_{\odot}$} & 6140.0/5718 & 2.2\\
\textit{vapec} & 5.0$_{-3.0} ^{+1.3}$ & 248$_{-12} ^{+16}$ & --- & 5.5$\pm0.5$ & 0.20$_{-0.02} ^{+0.05}$ & 0.21$_{-0.05} ^{+0.04}$ & 6145.1/5717 & 2.7 \\
\textit{vpshock} & 0 & 256$_{-6} ^{+18}$ & 47.5 ($>4.3$) & 2.34$_{-0.35}^{+1.18}$ & 0.46& 0.63 & 6160.2/5719 & 1.8\\
\noalign{\smallskip}
\tableline
 \end{tabular}

 \tablecomments{All errors are given at the 90\,\% confidence level. Parameters without errors were fixed in the fit.}
 \tablenotetext{\dag}{Emission measure $\int n_e n_H dV$}
 \tablenotetext{\ddag}{Unabsorbed X-ray luminosity in the 0.2--5~keV range.}
\end{table*}

\end{document}